\documentclass[conference]{sig-alternate-10pt}
\usepackage{url}
\usepackage{color}
\usepackage{multirow}

\usepackage{url}
\definecolor{gray}{gray}{0.5}

\usepackage{graphicx}%
\usepackage{dcolumn}%
\usepackage{bm}%
\usepackage{epsfig}
\usepackage{psfrag}
\usepackage{subfigure}
\usepackage{color}
\usepackage{amsmath, amssymb, latexsym,amsfonts,epsfig, graphicx, verbatim}
\usepackage[nospace]{cite}
\usepackage{algorithmic, algorithm}
\usepackage{verbatim}
\usepackage{xspace}

\newtheorem{theorem}{Theorem}[section]

\newcommand{\sss}[1]{{\scriptscriptstyle\textrm{#1}}}

\newcommand{\Mean}[1]{\mathbb{E}\left[#1\right]}

\newcommand{\eqn}[1]{Eq.(\ref{#1})}
\newcommand{\Sec}[1]{Section~\ref{#1}}
\newcommand{\Fig}[1]{Fig.~\ref{#1}}

\newcommand{\eq}{\!\!=\!}

\newcommand{\vol}{\textrm{vol}}
\newcommand{\w}{\textrm{w}}

\newcommand{\heur}{S-WRW\xspace}

\newcommand{\NRMSE}{{\small\textrm{NRMSE}\xspace}}

\newcommand{\inv}{{\scriptscriptstyle\textrm{\--1}}}

\newcommand{\C}{{\scriptscriptstyle\textrm{C}}}

\newcommand{\athina}[1]{[\textcolor{blue}{\textit{#1 (Athina)}}]}
\newcommand{\est}[1]{\widehat{#1}}

\newcommand{\ie}{{\em i.e., }}
\newcommand{\eg}{{\em e.g., }}

\begin{document}

\title{Coarse-Grained Topology Estimation via Graph Sampling\thanks{We make our datasets available, together with a customizable web-based visualization at {\tt www.geosocialmap.com}}}

\numberofauthors{6}

\author{
\alignauthor
Maciej Kurant\\%\titlenote{test}\\
       \affaddr{CalIT2}\\
       \affaddr{UC Irvine}\\
       \email{mkurant@uci.edu}
\and
\alignauthor
Minas Gjoka\\
       \affaddr{CalIT2}\\
       \affaddr{UC Irvine}\\
       \email{mgjoka@uci.edu}
\and
\alignauthor
Yan Wang\\
       \affaddr{CalIT2}\\
       \affaddr{UC Irvine}\\
       \email{wang.yan@uci.edu}
\and
\alignauthor
Zack W. Almquist\\
       \affaddr{Sociology Dept, CalIT2}\\
       \affaddr{UC Irvine}\\
       \email{almquist@uci.edu}
\and
\alignauthor Carter T. Butts\\
       \affaddr{Sociology Dept, CalIT2, IMBS}\\
       \affaddr{UC Irvine}\\
       \email{buttsc@uci.edu}
\and  %
\alignauthor Athina Markopoulou\\
       \affaddr{EECS, CalIT2, CPCC}\\
       \affaddr{UC Irvine}\\
       \email{athina@uci.edu}
}

\maketitle

\begin{abstract}

Many online networks are measured and studied via sampling techniques, which typically collect a relatively small fraction of nodes and their associated edges. Past work in this area has primarily focused on obtaining a representative sample of nodes and on efficient estimation of local graph properties (such as node degree distribution or any node attribute)  based on that sample. However, less is known about estimating the global topology of the underlying graph.

In this paper, we show how to efficiently estimate the coarse-grained topology of a graph from a probability sample of nodes.  In particular, we consider that nodes are partitioned into {\em categories}  (\eg countries or work/study places in OSNs), which naturally defines a weighted \emph{category graph}.  We are interested in estimating (i)~the size of categories and (ii)~the probability that nodes from two different categories are connected. For each of the above, we develop a family of estimators for design-based inference under uniform or non-uniform sampling, employing either of two measurement strategies: {\em induced subgraph sampling}, which relies only on information about the sampled nodes; and {\em star sampling}, which also exploits category information about the neighbors of sampled nodes. 
We prove consistency of these estimators  and  evaluate their efficiency via simulation on fully known graphs. We also apply our methodology to a sample of Facebook users to obtain a number of category graphs, such as the college friendship graph and the country friendship graph; we share and visualize the resulting data at \url{www.geosocialmap.com}.
\end{abstract}


\begin{figure}
\psfrag{A}[l][c][1]{Original graph $G$}
\psfrag{B}[l][c][1]{Category graph $G^\C$}
\psfrag{1}[c][c][0.8]{$\w(\circ,\bullet) =  \frac{3}{9}$}
\psfrag{2}[c][c][0.8]{$\w(\textcolor{gray}{\bullet}, \bullet)=\frac{1}{6}$}
\psfrag{3}[c][c][0.8]{$\w(\textcolor{gray}{\bullet}, \circ)=\frac{4}{6}$}
\includegraphics[width=0.47\textwidth]{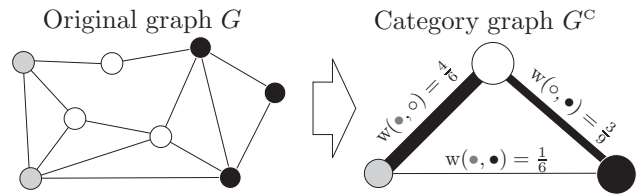}
\vspace{-0.3cm}
\caption{Nodes in the original graph ($G$) belong in one of three categories: white, gray, and black. The category graph ($G^\C$) consists of three nodes, corresponding to the three categories, connected by weighted edges.  The edge weight $\w(\circ,\bullet)$ in~$G^\C$ is the probability that a black and a white node, randomly chosen from~$G$, are connected in~$G$ (see~\eqn{eq:w_AB_basic2}).  The main goal of this paper is to estimate these edge weights based on a  probability sample of nodes of $G$.}
\vspace{-0.5cm}
\label{fig:category_graph}
\end{figure}

\section{Introduction}
Many large online networks, such as online social networks (OSNs) and the World Wide Web (WWW), are currently studied via sampling techniques. 
Sampling becomes necessary due to the sheer size of these networks and/or access limitations, which make it infeasible to collect (and, in some cases, to analyze) these networks in their entirety.

Most principled graph sampling methods to date have focused on collecting a %probability sample %
probability sample of nodes~\cite{Henzinger2000,Stutzbach2006-unbiased-p2p,Rasti09-RDS,Gjoka2010,Ribeiro2010,Gjoka2011_multigraph_JSAC,Ribeiro2010a,Avrachenkov2010,Kurant2011_SWRW}.
Based on such a sample, one can efficiently estimate many local graph properties, such as node attribute frequency, degree distribution, degree-degree correlations, or clustering coefficients~\cite{Hardiman2009,Kolaczyk2009}. 
However, these features reveal little about the global properties of the underlying graph, such as path-based properties (connectivity, diameter, average shortest path length) or community structure. 
%

%
%
%
%
%
\begin{comment}
An alternative approach, often taken in practice, is to collect nodes in the neighborhood of some (one or more) initial nodes, by using Breadth-First Search (BFS), Forest Fire, or similar techniques~\cite{Leskovec2006_sampling_from_large_graphs,Ahn-WWW-07,Mislove2007,MisloveWosn08}. 
The resulting sample is typically well-connected, and considerably denser than the graph induced on a uniform sample of nodes, but is also
 (i)~strongly dependent on the starting nodes and 
(ii)~subject to biases that cannot be systematically corrected in arbitrary graphs~\cite{Najork01,Lee-Phys-Rev-06,snowball-bias,Ye2010,Gjoka2010,Kurant2010}.  As a result, analyses based on such samples can prove misleading.
%
\end{comment}
%
%
%
In this paper, we show how a particular aspect of global network structure, 
namely coarse-grained topology, can be efficiently estimated from a probability sample of nodes.  
Specifically, we note that nodes in many online graphs belong to \emph{categories}, explicitly declared by users or clearly determined by observable characteristics.   
For example, in Facebook, users can officially declare the college or workplace with which they are affiliated, or a country/city in which they live. Similarly, in the WWW, all nodes can be categorized by their domain names, and the users of Internet radio sites like {\tt Last.FM} may be grouped on the basis of listening behavior. 
This potentially allows us to build and study \emph{category graphs},
in which each node corresponds to a category and edge weights reflect the frequency of ties between category members in the original graph. 
We illustrate these concepts in~\Fig{fig:category_graph}. 

The contribution of this paper lies in developing and evaluating several efficient estimators for two properties of the category graph, namely the size of the categories and the edge weights. 
These estimators take as input a uniform or non-uniform probability sample of nodes, measured via one of two strategies: {\em induced subgraph sampling}, in which we have information regarding only the sampled nodes; and {\em star sampling}, in which we also have category information about the neighbors of sampled nodes. 
We show that our estimators have good asymptotic properties (consistency, and hence asymptotic unbiasedness) and we evaluate their efficiency via simulation: employing fully observed graphs from both synthetic and empirical sources, we examine how estimator performance varies with the properties of the underlying graph.  Finally, as a practical illustration of our approach, we apply our methodology to a sample of Facebook nodes
to estimate several Facebook category graphs, such as the inter-college and inter-country friendship graphs.   The resulting Facebook category graphs are made available (along with a highly-customizable, web-based visualization service) at \url{www.geosocialmap.com}.

The structure of the remainder of the paper is as follows. 
Section~2 presents the problem statement. 
Section~3 reviews node sampling techniques. 
Sections~4 and~5 present our estimators for uniform and non-uniform probability samples, respectively.  Section~6 presents simulation results on fully known graphs. 
Section~7 applies our estimators to samples of Facebook. 
Section~8 reviews related work. Section~9 concludes the paper. 
Finally, in Appendix we prove the consistency of all estimators proposed in this paper.

\section{Notation and Problem Statement}

\subsection{Basic graph $G$}
We consider an undirected, static\footnote{Sampling dynamic graphs is currently an active research area~\cite{Stutzbach2006-unbiased-p2p, Rasti09-RDS,Willinger09-OSN_Research}, but out of the scope of this paper. 
Indeed, during the collection of Facebook data sets we use, the underlying graphs changed very insignificantly~\cite{Gjoka2010,Kurant2011_SWRW}.
Moreover, in this paper we focus on coarse granularity, which should change even more slowly in time, as argued in~\cite{Willinger09-OSN_Research}.
}
graph $G=(V,E)$, with $N\eq|V|$~nodes and $|E|$~edges.
Denote by $\deg(v)$ the degree of node $v\in V$, and by
\begin{equation}\label{eq:volume}
\vol(A) = \sum_{v\in A} \deg(v)
\end{equation}
the volume of a set of nodes $A\subseteq V$. We will often use
\begin{equation}\label{eq:relative fractions}
f_A = \frac{|A|}{|V|}  \quad \textrm{ and } \quad f^\sss{vol}_A = \frac{\vol(A)}{\vol(V)}
\end{equation}
to denote the relative size of $A$ in terms of number of nodes and volume, respectively.

\subsection{Category graph $G^\C$}

We assume that the set of nodes $V$ is partitioned into a set $\mathcal{C}$ of \emph{categories}, \ie that $\bigcup_{C\in \mathcal{C}}\eq V$.
We are interested in the \emph{category graph} $G^\C=(\mathcal{C}, E^\C)$, with node set given by the categories of~$G$.\footnote{We are not the first ones to be interested in coarse-grained structures.  See, \eg the social network literature on \emph{blockmodels} \cite{wasserman.faust:bk:1994}, in which our categories correspond to \emph{positions}, our category graph to the \emph{reduced graph} or \emph{block image}, and our edge weights to \emph{block densities} or \emph{mixing rates}.  
See \Sec{sec:related} for additional references. 
}
For two different categories $A,B\in\mathcal{C}$, $A\neq B$, denote by $E_{A,B} \subset E$ the corresponding edge-cut in $G$, \ie
$$E_{A,B}\ = \{\{u,v\}\in E:\ u\in A \textrm{ and }  v\in B\}.$$
If $|E_{A,B}|>0$ then we draw an edge $\{A,B\}$ between $A$ and $B$ in $G^\C$. 
We show an example of a category graph in~\Fig{fig:category_graph}. 

The way we defined category graph~$G^\C$ so far, prevents self-loops, but potentially allows for edge weights. 
The \emph{weight} $\w(A,B)$ of edge $\{A,B\}$ can be defined in a number of ways. 
For instance, one could trivially set it always equal to~1. 
In some settings, \eg statistical modeling,  %\maciej{What would be an example of that?}) %
the number of inter-category edges, $\w(A,B)\eq|E_{A,B}|$, is a useful choice.  
 For many purposes, however, it is useful to have a notion of edge weight that adjusts for category size, \eg
\begin{equation}
\label{eq:w_AB_basic2}
	\w(A,B)\ =\ \frac{|E_{A,B}|}{|A|\cdot|B|}.   %
\end{equation}
This definition has an intuitive interpretation.
Because $|A|\cdot|B|$ is the size of the maximum possible edge-cut from $A$ to $B$,  
 $\w(A,B)$ is equal to the probability that a uniformly selected member of~$A$ is connected to a uniformly selected member of~$B$.  
We give an example of these weights $\w(A,B)$ in \Fig{fig:category_graph}.

\subsection{Goal: Estimate $G^\C$ through sampling}
Given the full knowledge of graph~$G$, it is trivial to construct the category graph with its edge weights. %
In many cases, however, the knowledge of the full graph~$G$ is not available, rendering exact computation of \eqn{eq:w_AB_basic2} infeasible.
For instance, downloading the entire Facebook social graph via HTML scraping would require downloading and processing about 50 terabytes of HTML traffic~\cite{Gjoka2010}, which is rather prohibitive in practice.   

In contrast, it is often possible to collect a \emph{sample} $S\subseteq V$ of nodes of~$G$.   Note that we permit $S$ to contain multiple copies of the same node, \ie the sampling with replacement.  
The challenge, then, and the main goal of this paper is to estimate the category graph $G^\C$ based on the sample%
~$S$.

\section{Sampling}\label{sec:Sampling techniques}

Our methodology takes as input a probability sample of nodes. 
Obtaining such a sample is an active research topic in its own right (see~\Sec{sec:related}). 
In~\Sec{sec:node selection}, we briefly review the node sampling techniques that we use later in simulations and Facebook implementation.

Independently of the sampling technique employed, we may collect less or more category information on each sampled node. 
In~\Sec{sec:observation}, we describe two scenarios most common in practice. 
As we will see later, they result in two different sets of estimators, often with very different performance.

\begin{comment}
The basic graph $G$ consists of nodes and edges; furthermore, each node is associated with a category. During sampling, we select a subset of nodes, edges and category information, which we late use to estimate the category graph $G^\C$. This section describes our graph sampling designs, which can best be described by two distinct stages. In the first stage, we {\em select} nodes to include in the sample, according to a sampling probability. In the second stage,  we {\em observe} information associated with the selected nodes, specifically (i) some or all of the edges incident to that node and (ii) category information about the node and/or its neighbors. 
Section  \Sec{sec:node selection} and  \Sec{sec:observation} describe ways to perform these two stages, respectively.
%
%
\end{comment}

%
\subsection{Node sampling techniques}\label{sec:node selection}
 \vspace{-0.3cm}

\subsubsection{Independence Sampling}

Under independence sampling, we sample nodes independently from the set~$V$, with replacement. 
We distinguish two general cases: \emph{Uniform Independence Sampling (UIS)}, where sampling probabilities are uniform (the same for all nodes); 
and \emph{Weighted Independence Sampling (WIS)}, which samples $v$ with probability proportional to a known weight~$\w(v)$.

In general, UIS and WIS are not feasible in online networks because of the lack of sampling frame. For example, the list of all user IDs may not be publicly available, or the user ID space may be too sparsely allocated to permit rejection sampling.
Nevertheless, these techniques can occasionally be employed, either when permitted by fortuitous circumstances (see \eg use by~\cite{Gjoka2010,Gjoka2011_multigraph_JSAC}) or when deliberately ``down-sampling'' a large graph to speed analysis. 
Independence samplers are also conceptually important as a baseline for comparison with crawling-based sampling methods.

\subsubsection{Sampling via Crawling}\label{subsec:walks}

In contrast to independence sampling, crawling techniques are feasible in many online networks, and are thus the main focus of this paper.  %
The crawling methods described here lead to an approximate probability sample (asymptotically approaching UIS or WIS) from the node set, in the limit of increasing sample size.

\emph{Simple Random Walk (RW)}~\cite{Lovasz93} selects the next-hop node~$v$ uniformly at random among the neighbors of the current node~$u$. 
On a connected and aperiodic graph, RW samples node $v$ with probability linearly proportional to its degree~$\deg(v)$. 

\emph{Weighted Random Walk (WRW)} is RW on a weighted graph~\cite{AldousBookInPreparation}. 
In our simulations and implementation, we use ``Stratified WRW,'' or \heur~\cite{Kurant2011_SWRW}, \ie a version of WRW that increases the sampling efficiency by over-sampling graph regions relevant to the measurement objective and under-sampling the irrelevant ones.

\emph{Metropolis-Hastings Random Walk (MHRW)} is a version of random walk that 
modifies the transition probabilities to converge to a desired stationary distribution (often uniform). 
It was shown in~\cite{Rasti09-RDS, Gjoka2010} that RW 
outperforms MHRW for most applications, which we observe in our implementation as well.

\begin{figure}
\psfrag{A}[l][c][1]{\small (a)~Induced subgraph sampling}
\psfrag{B}[l][c][1]{\small \qquad (b)~Star sampling}
\psfrag{L3}[l][c][0.8]{Sampled nodes}
\psfrag{L2}[l][c][0.8]{Unsampled nodes, with known category}
\psfrag{L0}[l][c][0.8]{Unknown nodes}
\psfrag{L4}[l][c][0.8]{Observed edges}
\includegraphics[width=0.47\textwidth]{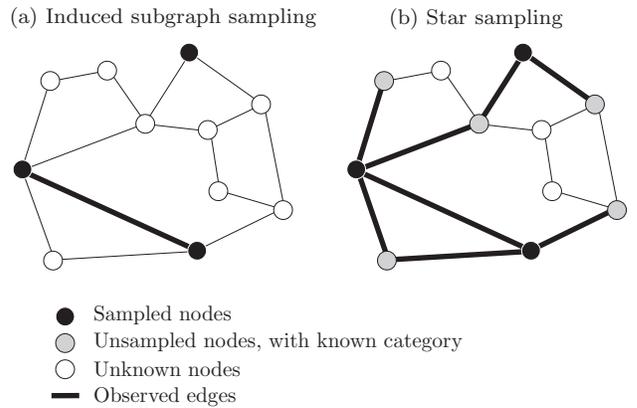}\vspace{-0.5cm}
\caption{Observed categories and edges, under two scenarios we study in this paper.}
\vspace{-0.4cm}
\label{fig:sampling_techniques}
\end{figure}

\subsection{Observed categories and edges}\label{sec:observation}

Our estimators will make use of every \emph{fully observed edge}, \ie edge $\{u,v\}$ for which we know the categories of both $u$ and~$v$. 
We distinguish between two measurement scenarios~\cite{Kolaczyk2009} that yield different sets of observed edges, as follows. %

\subsubsection{Induced Subgraph Sampling}\label{subsec:Induced Sampling}
Under \emph{induced subgraph sampling}, 
we learn the categories of the sampled nodes only. 
Consequently, the observed edges are only the edges induced on the set~$S$ of sampled nodes, as shown in~\Fig{fig:sampling_techniques}(a).

\subsubsection{Star Sampling}
In some settings, sampling a node $u\in S$ reveals the categories of \emph{all} its neighbors (not only the neighbors in~$S$). 
This is typically the case when sampling is done through scraping the HTML pages of OSNs~\cite{Gjoka2010,Kurant2011_SWRW}.
We refer to this as  \emph{star sampling}\footnote{To be precise, following the terminology of~\cite{Kolaczyk2009},   \emph{labeled} star sampling. The \emph{unlabeled} star sampling gets only the total number of neighbors, without their identities or categories.} and we show an example in \Fig{fig:sampling_techniques}(b). 

Finally, we emphasize that star sampling requires only information about neighbors' \emph{categories}; 
their degree or friend list is not needed, nor ties among neighbors (as in complete egonet sampling \cite{wasserman.faust:bk:1994}). 

\section{Uniform Sampling}  %
\label{sec:Uniform sampling}

In this section, we provide design-based estimators for category sizes and category graph edge weights, given a uniform independence (UIS) sample from the node set.  All estimators shown in this section and in \Sec{sec:Weighted sampling} are consistent; %
proofs are provided in the Appendix.

\subsection{Estimating category size ($|A|$)}\label{subsec:category size}
Learning the size of a given category can be an important measurement objective per se. 
Moreover, it is also a building block of the edge weight estimators we derive in~\Sec{subsec:UIS:Category edge weights:star}.

\subsubsection{Induced subgraph sampling}
The size $|A|$ of category $A$ can be trivially estimated by multiplying by $N$ the fraction of nodes sampled in $A$,~\ie
\begin{equation}\label{eq:category_size_induced}
    |\est{A}|\ =\ N\cdot\frac{|S_A|}{|S|},
\end{equation}
where 
$$S_A = \{v \in S : v\in A\}$$
is a multiset containing all samples 
from category $A$.

\subsubsection{Star sampling}\label{subsec:size_UIS_star}
Although not obvious at first blush, star sampling gives us an alternative way to estimate category sizes. 
Denote by 
$$k_A \ =\ \frac{1}{|A|} \sum_{v \in A} \deg(v)  \quad \textrm{and} \quad  k_V \ =\ \frac{1}{|V|} \sum_{v \in V} \deg(v)$$
the average node degree in category~$A$ and in the entire graph, $G$, respectively. 
Because $\vol(A)=|A|\cdot k_A$, we can re-write the relative volume $f^\sss{vol}_A$ of category~$A$ (see~\eqn{eq:relative fractions}) as 
$$f^\sss{vol}_A \ =\ \frac{\vol(A)}{\vol(V)} \ =\  \frac{|A| \cdot k_A}{|V| \cdot k_V} \ =\  \frac{|A| \cdot k_A}{N \cdot k_V}.$$
This allows us to estimate the size $|A|$ of category $A$ as 
\begin{equation}\label{eq:category_size_star}
    |\est{A}|\ =\ N\cdot \est{f}^\sss{vol}_A \cdot \frac{\est{k}_V}{\est{k}_A}.
\end{equation}
This formula may seem less attractive than~\eqn{eq:category_size_induced}, because we now have to estimate three different numbers. 
However, $k_V$ and $k_A$ can be easily estimated, respectively by
\begin{equation}
\label{eq:av node degree}
	\est{k}_V = \frac{\sum_{v \in S} \deg(v)}{|S|} \quad \textrm{and} \quad  \est{k}_A = \frac{ \sum_{v \in S_A} \deg(v)}{|S_A|}.
\end{equation}
Similarly, $f^\sss{vol}_A$ could be estimated by 
$$	\est{f}^\sss{vol}_A\ = \ \frac{\sum_{v \in S} \deg(v)\cdot 1_{\{v\in A\}}}{\sum_{v \in S} \deg(v)}.$$
But we have proposed in \cite{Kurant2011_SWRW} a much more efficient star-based estimator of $f^\sss{vol}_A$, \ie
\begin{equation}\label{eq:vol c 2 UIS}
	\est{f}^\sss{vol}_A \ \ =\  \ \frac{1}{\vol(S)}  \sum_{s\in S}\sum_{v \in \mathcal{N}(s)}\!\! 1_{\{v\in A\}}. 
\end{equation}
By plugging \eqn{eq:av node degree} and \eqn{eq:vol c 2 UIS} into \eqn{eq:category_size_star}, we obtain a complex yet powerful star-based estimator of size~$|A|$.

We show later that the star sampling estimator of \eqn{eq:category_size_star} often outperforms the trivial estimator or \eqn{eq:category_size_induced}, especially in dense graphs. 
One reason for this result is that \eqn{eq:category_size_induced} employs only the number $|S_A|$ of samples from $A$. 
This number is a random variable with a potentially high variance (especially for walks). In contrast, \eqn{eq:category_size_star} 
relies on mean degree estimates rather than on counting-based estimates, which employ more information (edges not in $G[S]$) and tend to be more stable.  

\subsection{Estimating category edge weights ($\w(A,B)$)}
Recall from \eqn{eq:w_AB_basic2} that, given the full knowledge of graph~$G$, the weight $\w(A,B)$ is obtained by dividing the number of edges between $A$ and $B$ by the maximal possible number of such edges. 
We use this same idea when estimating $\w(A,B)$ from our sample~$S$, except that now we divide the number of edges \emph{observed} between $A$ and $B$ by the maximal number of such edges we could potentially observe.

\smallskip
\subsubsection{Induced subgraph sampling}
Under induced subgraph sampling, we observe edges between the sampled nodes only. 
Consequently, in our sample we observe $\sum_{a\in S_A}\sum_{b\in S_B} 1_{\{\{a,b\}\in E\}}$ edges between distinct categories $A$ and $B$, out of the maximal number $|S_A|\cdot|S_B|$ we could possibly observe, leading to the trivial estimator
\begin{equation}\label{eq:w_AB_Both end-nodes sampled}
    \est{\w}(A,B)\ =\ \frac{\displaystyle \sum_{a\in S_A}\sum_{b\in S_B} 1_{\{\{a,b\}\in E\}} }{|S_A|\cdot|S_B|}.
\end{equation}
(Note that when $S$ contains the same node multiple times, we count any corresponding sampled edges multiple times as well.)

\subsubsection{Star sampling}\label{subsec:UIS:Category edge weights:star}
Under star sampling, on sampling node $a\in A$ we observe the set $E_{a,B} \subset E$ of all edges between~$a$ and category~$B\neq A$. 
So we observe $|E_{a,B}|$ edges out of a potential $|B|$ edges between~$a$ and~$B$. 
If we consider all nodes $S_A$ we sampled from~$A$, we observe $\sum_{a\in S_A} |E_{a,B}|$ out of a potential $|S_A|\cdot|B|$ edges. 
The same applies to nodes $S_B$ sampled in $B$ and their neighbors in $A$. 
Consequently, we can estimate the category graph edge weight $\w(A,B)$ by dividing the total number of edges we observed between $A$ and $B$ by our estimate of the maximal number we could potentially observe, \ie
\begin{equation}\label{eq:w_AB_star}
	\est{\w}(A,B)\ =\ \frac{\displaystyle \sum_{a\in S_A} |E_{a,B}| \ +\  \sum_{b\in S_B} |E_{b,A}|}
	{\displaystyle |S_A|\cdot|\est{B}|\ +\ |S_B|\cdot|\est{A}|}.
\end{equation}
Note that because we usually do not know the real sizes of $A$ and~$B$,~\eqn{eq:w_AB_star} uses their estimators $|\est{A}|$ and $|\est{B}|$. We can employ either~\eqn{eq:category_size_induced} or~\eqn{eq:category_size_star}, as needed.

Observe that the star sampling estimator is potentially more efficient than the trivial induced subgraph estimator, because we include edges (and non-edges) between sampled members of $A$ and $B$ and members of the respective sets that were \emph{not} themselves sampled.  For categories with large mean degree, this may represent a substantial increase in information versus the induced subgraph case.

\subsection{Population size~($N$)}
\label{subsec:Population size N UIS}

In our estimation of category sizes, the population size~$N\eq |V|$ is required. In some cases $N$ is known (e.g., in an OSN context, it may be published by the service provider), but in general this is not the case. 
Fortunately, where $N$ is not available, we can turn to estimation. For instance, \cite{Katzir2011} proposes an approach based on a ``reversed coupon collector'' problem, which can be used with both uniform and non-uniform sampling. 

Finally, we note that $N$ is only necessary where absolute values of category sizes are required.  Specifically, all edge weights and category sizes can be estimated up to a constant of proportionality without knowing the size of the total population.  Thus, if we are interested in ratios of category sizes and/or edge weights (\eg the relative weight of the $A,B$ connection versus the $A,C$ connection in $G^C$), then $N$ can be ignored (and replaced by an arbitrary constant in the above equations).

\section{Non-Uniform Sampling} %
\label{sec:Weighted sampling}
The estimators derived in \Sec{sec:Uniform sampling} hold under UIS, where every node $v\in V$ is sampled with the same probability. 
Such a sampling design is rarely feasible in practice. 
Moreover, in some cases UIS may be also undesirable, \eg when some categories are irrelevant to our measurement~\cite{Kurant2011_SWRW}.

A more common scenario is \emph{non-uniform} probability sampling, where every node $v\in V$ is sampled with probability proportional to a known weight~$\w(v)$. 
Indeed, this is the case for WIS, RW, \heur and other principled walk-based sampling methods, provided that samples have adequately converged \cite{Gjoka2010}.
Non-uniform samples are by definition biased towards nodes of higher weight (typically degree), which may dramatically distort the estimation results if used without correcting for sampling probabilities~\cite{Gjoka2011_Facebook_JSAC}. 

Fortunately, where sampling weights are known (as in the above designs), they can be corrected for by an appropriate (though not necessarily obvious) re-weighting of the measured values. 
In this section, we rewrite all estimators from~\Sec{sec:Uniform sampling} in such a corrected form.

\subsection{Correcting for sample bias}
\label{subsec:Correcting for the bias in RW, WRW and WIS}

A weighted sample can be unbiased using the Hansen-Hurwitz estimator~\cite{HansenHurwitz1943} as shown \eg in \cite{Salganik2004, VolzHeckathorn08} for random walks and also used in \cite{Rasti09-RDS}.  %
Let every node $v\in V$ carry a value $x(v)$. We can estimate the population total $x_\sss{tot} = \sum_v x(v)$ by
\begin{equation}\label{f_tot}
	\hat{x}_\sss{tot} = \frac{1}{n}\sum_{v\in S} \frac{x(v)}{\pi(v)},
\end{equation}
where $\pi(v)$ is the sampling probability of node~$v$. 

In practice, we usually know~$\pi(v)$, and thus~$\hat{x}_\sss{tot}$,  only up to a constant, \ie we know the (non-normalized) weights~$\w(v)$, $\w(v)\sim\pi(v)$. 
Fortunately, we can often address this problem by estimating the ratio of two totals, which makes the unknown constants cancel out. 
We will use this approach below.

\subsection{Estimating category size ($|A|$)}

\subsubsection{Induced subgraph sampling}

Following \eqn{f_tot}, we can estimate $|S_A|$ by setting $x(v)\equiv 1_{\{v\in A\}}$. This yields 
$|\est{S}_A| = \frac{1}{n}\sum_{v\in S} \frac{ 1_{\{v\in A\}}}{\pi(v)} \ = \ \frac{1}{n}\sum_{v\in S_A} \frac{ 1}{\pi(v)}.$
Analogously, $|\est{S}| = \frac{1}{n}\sum_{v\in S} \frac{ 1}{\pi(v)}$. Consequently, 
we can rewrite \eqn{eq:category_size_induced} as
\begin{eqnarray} 
\nonumber |\est{A}| &=& N\cdot\frac{\sum_{v\in S_A} \frac{ 1}{\pi(v)}}{\sum_{v\in S} \frac{ 1}{\pi(v)}} 
\ =\ N\cdot\frac{\sum_{v\in S_A} \frac{ 1}{\w(v)}}{\sum_{v\in S} \frac{ 1}{\w(v)}} \qquad {}\\
\label{eq:category_size_induced_weighted} &=& N\cdot\frac{\w_\inv(S_A)}{\w_\inv(S)},	
\end{eqnarray}
where 
$$
\w_\inv(X) = \sum_{v\in X}\frac{1}{\w(v)}
$$
is a `re-weighted size' of multiset $X\subset V$.

\subsubsection{Star sampling}

As in~\Sec{subsec:size_UIS_star}, we estimate the size of a category $A$ using \eqn{eq:category_size_star}, \ie
\begin{equation}\label{eq:category_size_star_WIS}
    |\est{A}|\ =\ N\cdot \est{f}^\sss{vol}_A \cdot \frac{\est{k}_V}{\est{k}_A}.
\end{equation}
However, now, the terms $\est{f}^\sss{vol}_A$, $\est{k}_V$ and $\est{k}_A$ must be calculated taking into account the sampling weights. 
Indeed, the weighted version of $\est{f}^\sss{vol}_A$ is (after~\cite{Kurant2011_SWRW})
\begin{equation}\label{vol c 2 WIS}
	\est{f}^\sss{vol}_A\ \ =\  \ \frac{1}{\displaystyle\sum_{s\in S} \frac{\deg(s)}{\w(s)}} \cdot  \sum_{s\in S} \left(\frac{1}{\w(s)}\sum_{v \in \mathcal{N}(s)}\!\! 1_{\{v\in A\}}\right).
\end{equation}
Similarly, the estimators \eqn{eq:av node degree} of $k_V$ and $k_A$ can be rewritten respectively by
\begin{equation}
	\label{eq:w_degree}
	\est{k}_V = \frac{\sum_{v \in S} \frac{\deg(v)}{\w(v)}}{\w_\inv(S)} \quad \textrm{and}\quad \est{k}_A = \frac{\sum_{v \in S_A} \frac{\deg(v)}{\w(v)}}{\w_\inv(S_A)}.
\end{equation}

\subsection{Estimating category edge weights ($\w(A,B)$)}

\smallskip
\subsubsection{Induced subgraph sampling}
Note that in the numerator of~\eqn{eq:w_AB_Both end-nodes sampled}, we have a sum over node \emph{pairs}, rather than single nodes. 
In this case, Hansen-Hurwitz estimator divides every component by the product 
of weights of the two involved nodes~\cite{Kolaczyk2009}, which yields
\begin{equation}\label{eq:w_AB_Both end-nodes sampled WIS}
    \est{\w}(A,B)\ =\ \frac{\displaystyle \sum_{a\in S_A}\sum_{b\in S_B} \frac{1_{\{\{a,b\}\in E\}}}{\w(a)\cdot\w(b)} }{\w_\inv(S_A)\cdot\w_\inv(S_B)}.
\end{equation}

\subsubsection{Star sampling}
Finally, under nonuniform sampling, \eqn{eq:w_AB_star} becomes
\begin{equation}\label{eq:w_AB_star_WIS}
	\est{\w}(A,B)\ =\ \frac{\displaystyle \sum_{a\in S_A} \frac{|E_{a,B}|}{\w(a)} \ +\  \sum_{b\in S_B} \frac{|E_{b,A}|}{\w(b)}}
	{\displaystyle \w_\inv(S_A)\cdot|\est{B}|\ +\ \w_\inv(S_B)\cdot|\est{A}|}.
\end{equation}
Again, we have two size estimators \eqn{eq:category_size_induced_weighted} and \eqn{eq:category_size_star_WIS} to choose from to plug into $|\est{A}|$ and $|\est{B}|$. We recommend selecting the one with smaller variance for the specific application. This variance can be estimated, \eg using bootstrapping \cite{efron.tibshirani:bk:1993}. 

\subsection{Sampling via crawling}

As we argued in \Sec{subsec:walks}, in many online networks the only feasible sampling approach is via crawling. 
Such techniques result in non-uniform sampling probabilities, and, consequently, sampling weights. 
For example, under RW the sampling weights converge asymptotically to $\w(v)\eq\deg(v)$~\cite{Lovasz93}.
Using these weights in conjunction with the WIS estimators above allows for consistent estimation of coarse-grained topology from random walk samples.

Of course, consecutive samples collected by crawls are in general correlated, which can potentially affect the efficiency of our estimators. 
One way to deal with that is to take, say, every $T$-th sample. 
For $T$ large enough, this \emph{thinning} technique effectively reduces sample correlations, at a cost of discarding a large portion of available information. 
Thinning is crucial in some applications, \eg those primarily based on counting repeated nodes, as in~\cite{Katzir2011}.  The ergodicity of standard random walk designs, however, guarantees convergence to the target (WIS) distribution with any effect of autocorrelation vanishing in the limit of sample size.  (See Appendix.)

\begin{figure*}

 \psfrag{rand=0.5,  category=8}[c][][0.8]{$\alpha=$}
 \psfrag{sample size}[c][][0.8]{sample size $|S|$}
 \psfrag{nrmse(size)}[c][][0.8]{$\NRMSE(|\est{A}|)$}
 \psfrag{nrmse(weight)}[c][][0.8]{$\NRMSE(\est{\w}(A,B))$}
 \psfrag{CDF}[c][][0.8]{CDF}
 \centering
 \includegraphics[width=.99\textwidth]{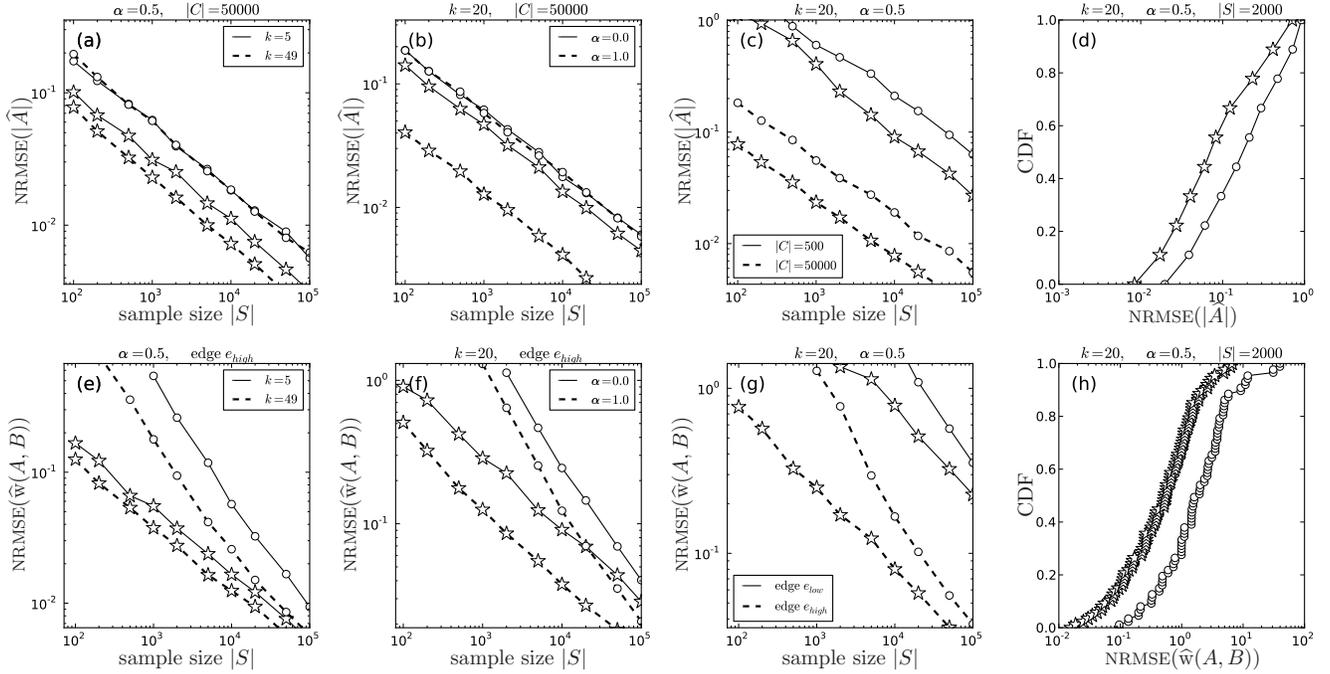}
 \caption{Simulations of UIS on synthetic graphs. 
 We estimate category sizes (top) and category edge weights (bottom),  using induced subgraph sampling (circles) and star sampling (stars). 
 } 
 \label{fig:simulations_category_graph} %
\end{figure*}

\section{Simulation Results}\label{sec:Simulation Results}

\subsection{Objective and performance metrics}

In this section, we apply our methodology to fully observed graphs from both synthetic and empirical sources.
Our objective is to evaluate estimator performance by comparing the (known) values of the category sizes and edge weights in each case with the values inferred using our methods.
We use the Normalized Root Mean Square Error (\NRMSE) to assess estimation error: %
\begin{equation}\label{eq:NRMSE}
\NRMSE(\est{x}) = \frac{ \sqrt{   \Mean{(\est{x}-x)^2 }    }}{x},
\end{equation}
where~$x$ is the real value 
 and $\est{x}$ is the estimate.

\subsection{Generated topologies} \label{subsec:Generated topologies}

First, we consider synthetic graphs. By simulating~$G$, we control many crucial parameters (such as graph density, or category size and tightness) and study the effect of these parameters on the efficiency of our estimators.

\subsubsection{Graph model}
We consider a graph $G$ with $N=88,850$ nodes partitioned into 10 categories. Their sizes range from $|C|\eq50$ to $|C|\eq50000$. 
Initially, nodes in each category form  a $k$-regular random graph, with the average node degree ranging from $k\eq5$ to $k\eq49$.
In addition, we add $N\cdot k /10$ random edges between nodes in different categories. The resulting graph~$G$ is connected (in all instances we used) and has $|E|=0.6\cdot N\cdot k$ edges. 
 By construction, $G$~has a very strong community structure. In order to study the effect of community tightness, we next  permute randomly the category labels of a fraction $\alpha\!\in\![0,1]$ of nodes. For $\alpha\eq 0$, node categories follow the strong community structure, whereas for $\alpha\eq 1$ the categories are completely independent of the graph structure.

\subsubsection{Category sizes}
We first study the efficiency of the category size estimators,  \eqn{eq:category_size_induced} and~\eqn{eq:category_size_star}. 
We present the results in the top row of~\Fig{fig:simulations_category_graph} and make the following observations.

In all of our simulated cases, all estimators converge to the true value as sample size increases.
Moreover, the star estimator performs better than the induced subgraph estimator, although its efficiency can depend on properties of $G$. 
For example, (i) the denser the graph, the better the star estimator is (\Fig{fig:simulations_category_graph}(a)), 
but (ii) its efficiency can be limited when clustering closely follows the category structure (\Fig{fig:simulations_category_graph}(b)). %
In contrast, the induced subgraph estimator is not affected by any of these properties. 
We also observe that both estimators perform better for larger categories (\Fig{fig:simulations_category_graph}(c)). 
In \Fig{fig:simulations_category_graph}(d), we show the CDF of the NMSE of all  (ten) estimators of the category sizes.

\subsubsection{Category edge weights}

In the bottom row of~\Fig{fig:simulations_category_graph}, we use \eqn{eq:w_AB_Both end-nodes sampled} and \eqn{eq:w_AB_star} to estimate the category edge weights under induced and star sampling designs, respectively. 

Again, both estimators converge, with the star estimator performing better than the induced one. 
As before, the star estimator benefits from higher graph density (\Fig{fig:simulations_category_graph}(e)) and looser category structure (\Fig{fig:simulations_category_graph}(f)). However, in this case the induced estimator is affected by these properties as well. 
Finally, in \Fig{fig:simulations_category_graph}(g) we compare the estimation efficiency of low-weight edge~$e_{low}$ (defined as the edge with $ 25^{th}$ percentile weight) with the high-weighted edge~$e_{high}$ ($75^{th}$ percentile).
As before, both estimators perform better for large estimated values.

\begin{table}[t!]
  \centering
  {\footnotesize
\begin{tabular}{|r|r|r|r|}
\hline
    Dataset          & $|V|$   & $|E|$ & $k_V$  \\%&  Description \\		
\hline
Facebook: Texas~\cite{Traud2011} &  36\,364 &    1\,590\,651  & 87.5   \\%&  Facebook Texas network~\cite{Traud2011}\\    
Facebook: New Orleans~\cite{Viswanath2009}&      63\,392 &     816\,885 & 25.8 \\%&  Facebook New Orleans network~\cite{Viswanath2009}\\    
      P2P~\cite{Leskovec2007} &      62\,561 &     147\,877 &  4.7 \\%&   Gnutella peer to peer network~\cite{Leskovec2007}\\
       Epinions~\cite{Richardson2003}&      75\,877 &     405\,738 & 10.7 \\%&  Who-trusts-whom network of Epinions.com \cite{Richardson2003}\\
\hline
\end{tabular}}
  \caption{Empirical topologies used in Sec.~\ref{subsec:Real-life topologies}. %
  }
  \label{tab:Real-life topologies}
  \vspace{-0.3cm}
\end{table}

\begin{figure*}

 \psfrag{size}[c][][0.9]{sample size $|S|$}
 \psfrag{median nrmse(size)}[c][][0.9]{median $\NRMSE(|\est{A}|)$}
 \psfrag{median nrmse(weight)}[c][][0.9]{median $\NRMSE(\est{\w}(A,B))$}
  
 \psfrag{CDF}[c][][0.8]{CDF}

 \centering
 \includegraphics[width=0.95\textwidth]{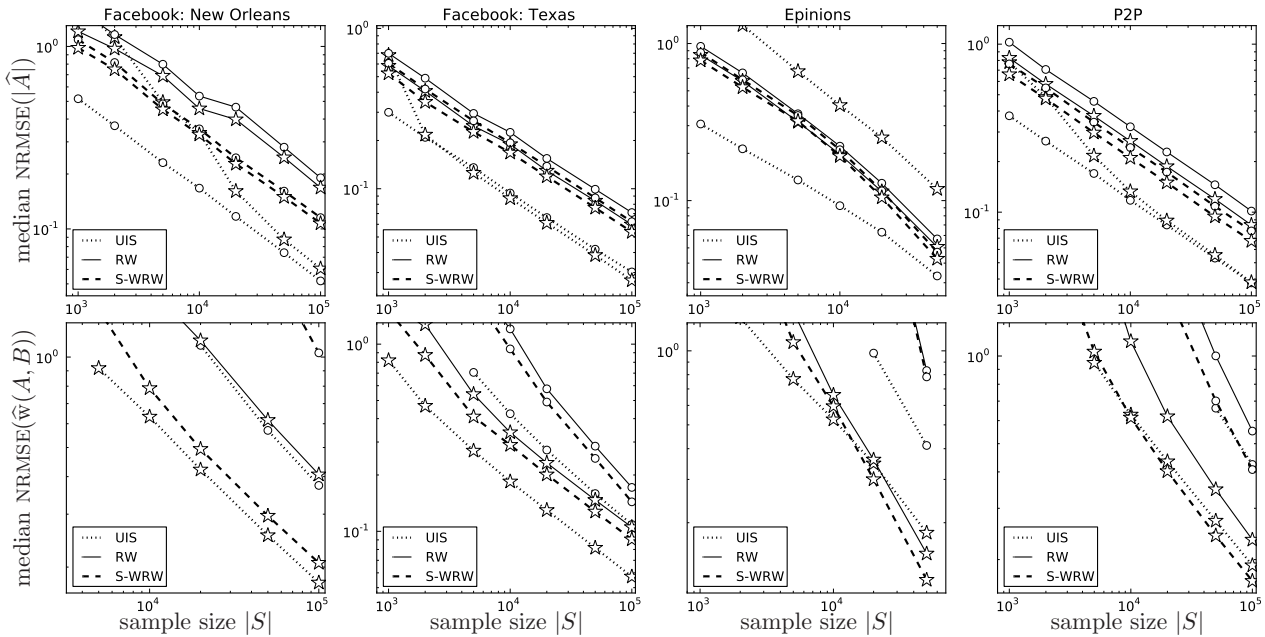}
 \caption{Simulations on empirically observed graphs. 
 We estimate category sizes (top) and category edge weights (bottom),  using induced subgraph sampling (circles) and star sampling (stars).}
 \label{fig:sim_real_graphs}
\end{figure*}

\subsection{Empirically observed topologies}\label{subsec:Real-life topologies}

\subsubsection{Datasets}

We consider four fully known topologies described in Table~\ref{tab:Real-life topologies}. 
We use two graphs extracted from Facebook because (i)~they significantly differ in density, and (ii)~Facebook is our focus in the experimental study of~\Sec{sec:Applications}.

In \Sec{subsec:Generated topologies}, we have seen that star sampling performs the worst if categories are aligned with the communities (dense clusters) existing in graphs. 
We decided to simulate presumably the worst-case category partition from the star sampling point of view. 
In particular, we use a standard community finding algorithm based on eigenvalues \cite{newman_eigen_06} to identify the 50 largest communities, and define each such community to be a category. All the remaining smaller categories (if any) are then grouped together as the $51^{st}$ category.

From these known graphs we then generate synthetic datasets by three different sampling methods: UIS, RW and \heur. 
Under \heur~\cite{Kurant2011_SWRW}, we use equal category weights for all categories, and we set $\tilde{f}_\ominus=0$ (because there are no irrelevant categories) and $\gamma=\infty$ (for simplicity).  As previously, our interest is in whether our estimators (applied to these realistic samples) will accurately reconstruct the true properties of the graphs in question.

\subsubsection{Category size}

We study the efficiency of the category size estimators in the top row of~\Fig{fig:sim_real_graphs}. 
Due to lack of space, we only report the median \NRMSE~across all categories. 
In \Fig{fig:simulations_category_graph}(d), this would correspond to the points on the horizontal line~$Y=0.5$.

The main observation is that, in contrast to \Sec{subsec:Generated topologies}, here the induced estimators can outperform the star estimators. 
This is particularly visible under UIS, probably because of the highly skewed node degree distributions. %
Such a distribution increases the variance of the average degree estimator~$\est{k}_A$ that is used in the star-based size estimation in~\eqn{eq:category_size_star}.%
\footnote{We might address this problem by modifying~\eqn{eq:category_size_star} to take \eg $\est{k}_A\eq\est{k}_V$ or a similar model-based extension. %
Such modifications may greatly reduce the variance of size estimation, albeit at the cost of some bias. 
(Indeed, this is an example of the classic ``precision vs accuracy'' tradeoff.)
Note that such modifications can allow us to use \eqn{eq:category_size_star} to estimate~$|A|$, even if none of our sampled vertices were drawn from~$A$.
Our initial experiments with such modifications have been encouraging, but we do not treat them in depth here.
}

However, in contrast to UIS, under RW and \heur star sampling usually performs better. This can also be explained by the highly skewed node degree distribution. 
Indeed, because both RW and \heur visit high-degree nodes more often than UIS, their star samples inherently collect and exploit more information about neighbor categories, which translates to a better performance.  
This effect is similar to the better star sampling performance under higher graph density in~\Sec{subsec:Generated topologies}.

\subsubsection{Category edge weights}

While there is no clear winner in the category size estimation, 
in the category edge weight estimation star sampling consistently and significantly outperforms induced sampling. 
Indeed, in \Fig{fig:sim_real_graphs}(e-h), the induced estimators often need 5-10 times more samples to achieve the same accuracy as star estimators.

UIS clearly performs best, especially when estimating category sizes. 
Not surprisingly, direct independence sampling should be preferred whenever available. 
In the more practical scenarios, however, we are limited to exploration-based techniques. 
In our simulations, \heur is consistently better than RW. 
Note that because all categories (and thus nodes) are relevant, this advantage of \heur is purely due to stratification. 
Moreover, the advantage of \heur increases with higher heterogeneity of category sizes (not shown here), which is in agreement with~\cite{Kurant2011_SWRW}.

\section{Facebook Category Graphs}\label{sec:Applications}

\noindent In this section, we use the estimators developed in this paper to infer several category graphs from Facebook.

\begin{table}[t!]
  \centering
  \footnotesize

\begin{tabular}{|@{}p{1.30cm}@{}|@{}l@{}|@{}c@{}|@{}c@{}|@{}l@{}|}

\hline
    Dataset     &  Studied categories   &  Crawl  &  \%  categ. & \# total       \\		
                &                       &   type  &  samples     &   samples      \\		
\hline

\multirow{3}{*}{2009 \cite{Gjoka2010} }       & \multirow{3}{3.1cm}{Regional (507) \\ \small{(34\% of population)} }     %
                       & MHRW09  &    34\%   & 28x81K      \\
      &                &  RW09   &    41\%   & 28x81K     \\
      &                &  UIS09  &    34\%   & 28x35K     \\

\hline
\hline
\multirow{2}{*}{2010 \cite{Kurant2011_SWRW} }&   \multirow{2}{3.1cm}{Colleges (10K+)\\ \small{(3.5\% of population)} } %
                                & RW10      &  9\%    & 25x40K      \\
     &                & S-WRW10   &  86\%   & 25x40K        \\

\hline
\end{tabular}
  \caption{Facebook datasets. }
  \label{tab:facebook_datasets}
  \vspace{-0.3cm}
\end{table}

\subsection{Data sets} \label{subsec:Data sets}

In our previous work~\cite{Gjoka2010,Kurant2011_SWRW}, we collected samples of Facebook users (about 10.1 million total users), with publicly available information. These datasets are summarized in Table~\ref{tab:facebook_datasets} and are used as input for the estimators of this paper. 
These datasets were collected using HTML scraping, which allowed us to collect for each user~$v$  not only $v$'s category, but also the list of $v$'s friends together with their categories; \ie we effectively collected a star sample of Facebook users. By discarding the information about $v$'s nodes, we can also use the induced subgraph estimators, for comparison.

\begin{figure}

 \psfrag{swrw10}[l][][0.55]{${}$\hspace{-0.99cm}S-WRW10} 
 \psfrag{rw2010}[l][][0.55]{${}$\hspace{-0.99cm}RW10} 
 \psfrag{uni2009}[l][][0.55]{${}$\hspace{-0.99cm}UIS09} 
 \psfrag{rw2009}[l][][0.55]{${}$\hspace{-0.94cm}RW09} 
 \psfrag{mhrw09}[l][][0.55]{${}$\hspace{-0.99cm}MHRW09} 
 
 \psfrag{samples}[c][][0.9]{\# Samples}  
 \psfrag{networks}[c][][0.9]{Facebook categories}  
 \centering
 \includegraphics[width=0.45\textwidth]{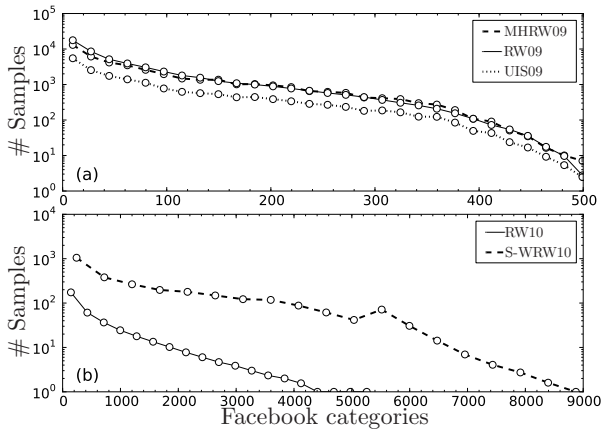}
 \caption{Number of samples per category in Facebook 2009 (top) and 2010 (bottom).}
 \label{fig:facebook_samples}
\end{figure}

{\bf The 2009 data sets:}
These data sets was collected in April 2009~\cite{Gjoka2010}, using three existing sampling techniques, UIS, MHRW and RW, as summarized in Table~\ref{tab:facebook_datasets}.
At that time, a Facebook user could be a member of any of four different types of categories, called ``networks'' in the Facebook terminology. 
Three of them, \emph{high school}, \emph{college} and \emph{workplace}, required passing a verification process, usually based on an email account from the institution in question. 
The fourth category, \emph{geographical region}, did not require any verification, and indicated the user's city, state or country. 
In this paper, we consider the geographical region categories from the 2009 data sets.
Each dataset consisted of 100-1000 samples from each of the 507 geographical regions, as shown in~\Fig{fig:facebook_samples}(a); UIS collected about two times fewer samples than the other two techniques.  
{\bf The 2010 data sets:}
The geographical region category was phased out in June 2009. 
Therefore, the data sets we collected in 2010~\cite{Kurant2011_SWRW} contain only the three remaining categories, from which we chose colleges as the category studied in this paper. Furthermore, Facebook  switched from 32 bit  to 64 bit userIDs, thus leading to a sparse userID space, which made UIS impractical to apply.  For this reason,
 in our 2010 Facebook data sets we collected only a RW sample (because RW proved to outperform MHRW~\cite{Rasti09-RDS,Gjoka2010}) as well as three variants of \heur~\cite{Kurant2011_SWRW}.  A full length (1M) RW typically collected only 0-10 samples of a particular college (\Fig{fig:facebook_samples}(b)). This is because of a relatively small college population (about $3.5\%$)  and a large number of colleges (more than $10,000$). 
Fortunately, \heur, a technique designed to oversample particular categories (here colleges), improves that result by at least one order of magnitude.

\begin{figure}

 \psfrag{size}[c][][0.9]{sample size $|S|$}
 \psfrag{median nrmse(size)}[c][][0.9]{median $\NRMSE(|\est{A}|)$}
 \psfrag{median nrmse(weight)}[c][][0.9]{median $\NRMSE(\est{\w}(A,B))$}

 \psfrag{swrw10}[l][][0.55]{${}$\hspace{-0.99cm}S-WRW10} 
 \psfrag{rw2010}[l][][0.55]{${}$\hspace{-0.99cm}RW10} 
 \psfrag{uni2009}[l][][0.55]{${}$\hspace{-0.99cm}UIS09} 
 \psfrag{rw2009}[l][][0.55]{${}$\hspace{-0.94cm}RW09} 
 \psfrag{mhrw09}[l][][0.55]{${}$\hspace{-0.99cm}MHRW09}

 \centering
 \includegraphics[width=0.5\textwidth]{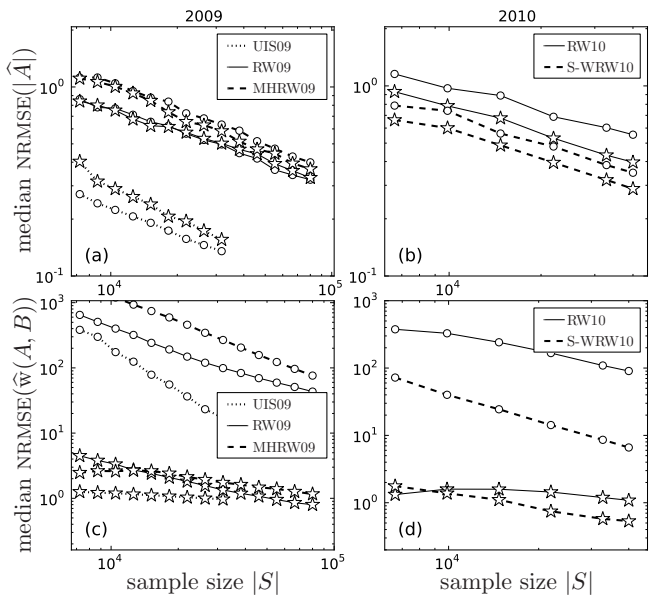}
 \caption{Results for 100 most popular regional networks in 2009 (a,c) and 100 college networks in 2010 (b,d) : category size estimation (a,b), and edge weight estimation (c,d).}
 \label{fig:facebook_node_edge_estimation}
\end{figure}

\subsection{Category graph estimation}\label{sec:Category Graph Estimation}

We present our results in~\Fig{fig:facebook_node_edge_estimation}. 
To calculate \NRMSE~we use as ground truth the average of estimation over all samples for each crawl type. 
In addition, we treat each of the 28 and 25 different walks, for the 2009  and 2010 data sets respectively, as a different sample.

\subsubsection{Category size}

We show the results of Facebook category size estimation in~\Fig{fig:facebook_node_edge_estimation}(a,b). 
Similarly to what we observed in the simulations in~\Sec{sec:Simulation Results}, UIS performs the best, and \heur outperforms RW. 
MHRW performs the worst, which was also expected given the recent studies of MHRW in~\cite{Rasti09-RDS, Gjoka2010}.  
Under UIS,  the induced estimator performs better.
Under RW and \heur, the star version is better, especially when categories are small, as in the 2010 data set.

\subsubsection{Category edge weight}

The estimation of category edge weights in Facebook, shown in~\Fig{fig:facebook_node_edge_estimation}(c,d), also confirms the observations in the simulations of~\Sec{sec:Simulation Results}. Indeed, all star estimators dramatically outperform their induced counterparts. 
And, as before, the sampling techniques ordered from the best to worst are: UIS, \heur, RW and MHRW.

\smallskip
Finally, note that $\NRMSE$s in \Fig{fig:facebook_node_edge_estimation}(a-d) are relatively high, even under star (\ie the better performing) sampling.
This is because these plots reach only relatively small sample sizes $|S|$ (\ie 25 or 28 times smaller than the entire sample at our disposal). 
Therefore, one could extrapolate the plots in~\Fig{fig:facebook_node_edge_estimation} by much more than a decade to the right, further reducing the values of $\NRMSE$.  Moreover, in the data sets that we eventually prepare, we combine together several outcomes of different, independent sampling techniques, which should further limit the estimation variance. 
Therefore, the results in~\Fig{fig:facebook_node_edge_estimation} should be treated as a guideline about the relative efficiency of the sampling techniques, rather than a comparison of the the absolute values of~\NRMSE.

 \begin{figure}[t!]
 \centering
\subfigure[Intra-continental country connections: 
Note the strong cliques formed between Middle Eastern countries and South-East-Asian countries.%%
There is no Facebook in China.]
{\includegraphics[width=0.45\textwidth]{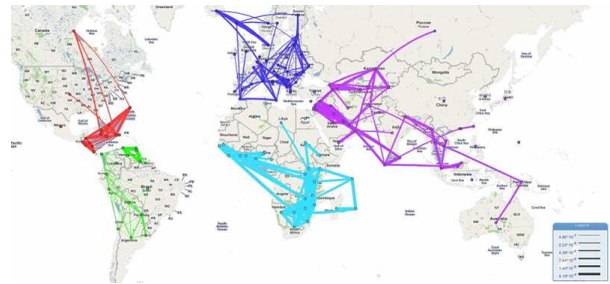}\label{fig:world}}
\subfigure[North-American regions: 
Physical distance is a major factor in the United States (red), but seemingly less so in Canada (green). Additionally, US and Canada are relatively weakly interconnected (thin blue lines).]%
{\includegraphics[width=0.45\textwidth]{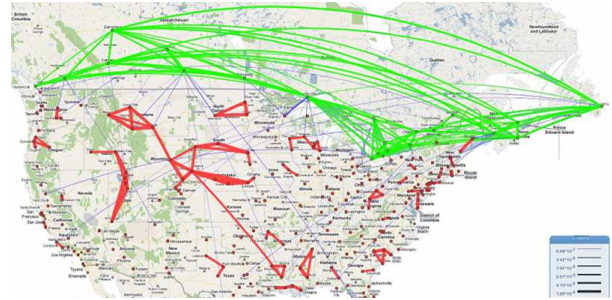}\label{fig:northmamerica}}
\subfigure[Top 133 US colleges according to the ``US News World Report'09'': 
Physical distance is a major factor for public colleges (green), but seemingly less so for private ones (red).]  %
{\includegraphics[width=0.45\textwidth]{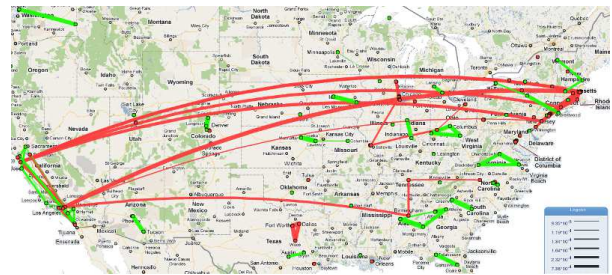}\label{fig:colleges}}
\caption{The friendship graph between regional networks. Available at \tt{www.geosocialmap.com}  }
\label{fig:screenshots}
\end{figure}

\subsection{Geosocial visualization}

Finally, we have developed a highly customizable, web-based tool for visualization of our Facebook category graphs. We have made a beta-version of the tool available at  \url{www.geosocialmap.com} and invite the reader to use it to experiment with the category-graphs described in this paper. This can be used to gain insight into the friendship relations among these categories, as defined {\em in Facebook}.\footnote{However, one should be careful about declaring categories in Facebook  as  representative of the real world. First, Facebook attracts some age groups more than others. Second, many Facebook users do not declare (or hide) their category membership. 
Finally, a user might have mistakingly chosen her category. For example, the third strongest link for ``Greece'' is ``Athens, GA``, which is clearly mistaken for Athens, Greece.}

\subsubsection{Cross-country friendships}\label{subsec:Country-level Friendships}

As mentioned earlier in \Sec{subsec:Data sets}, the 2009 data set contains the geographical region information, at various
granularities depending on Facebook's penetration in that region. This may may be either a user's city or state, (\eg for USA, Canada, UK) or the entire country (more typically). 

As an example, we create the country-to-country  friendship graph. To this end, we first merged together all categories coming from the same country. 
Next, we estimated the sizes of the resulting categories. Because, according to~\Fig{fig:facebook_node_edge_estimation}(a), the UIS induced sampling performed exceptionally well, we used it in the category size estimation. 
This information was next fed to the star estimators of category edge weights. 
Finally, for every edge, we take the average of the three estimates (resulting from UIS, MHRW and RW). %
\Fig{fig:screenshots}(a) presents a subset of ``The world according to Facebook'' graph.

\subsubsection{North America}
For the USA and Canada, the 2009 data set contains the geographical information at the granularity of 272 counties and provinces.  %
This allows us to create the North American friendship map. 
We followed  the same steps as in \Sec{subsec:Country-level Friendships}. 
An example is presented in \Fig{fig:screenshots}(b).

\subsubsection{US colleges}
Both the 2009 and 2010 data sets contain college categories. 
We chose the 2010 data set to create a college-to-college friendship graph. 
This data set consists of one RW sample and three \heur ones. 
Because \heur performed much better than RW (see~\Fig{fig:facebook_node_edge_estimation}(b,d)), 
we decided to use the three \heur samples only. 
Moreover, this time we estimated the size with the help of the star estimators, because they performed better~(\Fig{fig:facebook_node_edge_estimation}(b)).
Finally, as before, we fed the resulting category sizes into the star estimators of category edge weights, and we averaged the three \heur estimates into a final estimate.  
\Fig{fig:screenshots}(c) presents a subgraph of the resulting category graph.
%

\begin{comment}

\subsection{Discussion}
%
%
%

The above Facebook category graphs, and the customizable visualization in \url{www.geosocialmap.com}, can be used to better understand
the friendship relations of these categories, as defined {\em in Facebook}.

%
%

%
%
%
%
%

However, one should be careful about declaring categories 
in Facebook  as  representative of the real world. First, Facebook attracts some age groups more than others. 
Second, many users decide not to declare (or to hide) their membership in 
%
networks of Facebook, which might also introduce a bias. %
Third, a node $v\in A$ may have most of its friends in a network $B\neq A$ for many possible reasons. For example, $v$ might originate from $A$, but live in $B$; or setting $v\in A$ was a deliberate choice of $v$, to protect her privacy or just for a joke.
Finally, a user might have mistakingly chosen its network. For example, the third strongest link for members of the Greece network is ``Athens, GA``, most likely because they confuse it with the capital of Greece, Athens.

%

%

  %

\end{comment}

\section{Related work}\label{sec:related}

\smallskip\noindent\textbf{Node sampling in graphs.}
Most state-of-the-art crawling-based node sampling techniques use variants of random walks~(RW), 
such as the classic RW~\cite{Lovasz93,Heckathorn97_RDS_introduction,Salganik2004,Rasti09-RDS,Gjoka2010}, 
Metropolis-Hasting RW (MHRW)~\cite{Metropolis1953,mcmc-book,Stutzbach2006-unbiased-p2p,Rasti09-RDS,Gjoka2010}, 
multiple dependent RW~\cite{Ribeiro2010}, 
multigraph RW~\cite{Gjoka2011_multigraph_JSAC}, 
RW with jumps~\cite{Henzinger2000,Leskovec2006_sampling_from_large_graphs,Ribeiro2010a,Avrachenkov2010}, 
and weighted RW~\cite{Kurant2011_SWRW}. 
Based on the resulting (uniform or non-uniform) sample of nodes, there exist principled methods to estimate local graph properties (degree distribution, assortativity and clustering coefficient). \cite{Kolaczyk2009} is an excellent introduction; other examples include~\cite{Stumpf2005,Lee-Phys-Rev-06,Rasti09-RDS,Hardiman2009,Gjoka2010,Ribeiro2010,Ribeiro2010a,Avrachenkov2010,ahmed_win10,Gjoka2011_Facebook_JSAC}.
In our prior work \cite{Gjoka2010, Gjoka2011_multigraph_JSAC,Kurant2011_SWRW}, we used random-walk based crawls to collect user samples, which we use as input to the estimators proposed in this paper.

\smallskip\noindent\textbf{Topology inference.}
Much classic work on inference for basic network properties from node samples was done by Ove Frank and colleagues; see particularly \cite{frank:sjs:1977,frank:jspi:1977,frank:sn:1978}, which introduce Horvitz-Thompson estimators of edge totals (\ie volumes) from probability samples of nodes.  Early results involving topology inference from induced subgraph and star sampling were reviewed by \cite{frank:mish:1988}.  This prior work focused on the case of known population and category sizes, and assumed without-replacement designs.

Breadth First Search (BFS) has been used to sample topology \eg in~\cite{Ahn-WWW-07,Mislove2007,MisloveWosn08}.
However, a BFS sample is known to introduce a strong bias towards high degree nodes~\cite{Najork01,Lee-Phys-Rev-06,snowball-bias,Ye2010,Gjoka2010,Kurant2011_JSAC_BFS}, which makes it not representative with respect to many metrics. 
Although this degree bias can often be significantly corrected for~\cite{Kurant2011_JSAC_BFS}, 
the BFS sample covers only the neighborhood of the arbitrary starting node, which is not necessarily representative of the entire topology.  

\cite{Leskovec2006_sampling_from_large_graphs} evaluates a number of sampling methods and the graphs they induce. 
The authors conclude that Forest Fire~\cite{Leskovec05_Forest_Fire}, intuitively a hybrid of RW and BFS, produces topology samples that resemble the original graph the most. However, Forest Fire is subject to the same biases as BFS described above.

Another approach for inferring network structure is matrix completion of the distance matrix~\cite{Eriksson2011,Xu2011}. 
However, this approach faces its own challenges when applied to OSN samples. 
First, the distance matrix is typically high rank and one has to carefully identify a low rank structure~\cite{Eriksson2011}. 
Second, unlike traceroutes or tomographic techniques, crawling does not yield a random sample of distances~\cite{Eriksson2011,Xu2011}.

\smallskip\noindent\textbf{Induced subgraph vs. star sampling}
\cite{Kolaczyk2009} is a good summary of these two sampling designs. 
Induced subgraph sampling has been studied, \eg in \cite{Leskovec2006_sampling_from_large_graphs, Lee-Phys-Rev-06, Kolaczyk2009}
Star sampling is similar to egonet sampling \cite{wasserman.faust:bk:1994}, except that under star sampling we do not see edges between neighbors of a sampled node.  Our contribution here is to apply these measurement schemes in the context of category graph estimation.

\smallskip\noindent\textbf{Block models and mixing rates}
The use of partitions to produce reduced-form versions of larger networks has an extensive history in the social network literature, primarily under the label of ``block modeling;'' see \cite{wasserman.faust:bk:1994,nunkesser.sawitzki:ch:2005} for extensive reviews.  Block models with known partitions are sometimes called ``confirmatory'' block models, and have been studied largely from a statistical point of view (e.g., \cite{fienberg.wasserman:sm:1981,fienberg.et.al:jasa:1985}%
 and \cite{wasserman.faust:bk:1994} ch. 16).  Much of the latter interest is in modeling the edge weights (``block densities'' or ``mixing rates'') from covariates or other information in a fully-observed context, with considerable additional interest in the case where the network is observed but the categories are latent \cite{snijders.nowicki:joc:1997,nowicki.snijders:jasa:2001}.  Estimation of mixing rates from uniform node samples for categories of known size is also a well-known problem (see., e.g., \cite{frank:sjs:1977,yamaguchi:jasa:1990,morris:mb:1991}).  Comparable methods for link-trace samples are less well-developed, though see \cite{Heckathorn97_RDS_introduction,heckathorn:jsp:2002,handcock.gile:aas:2010,gile.handcock:sm:2010}.

Although estimation of mixing rates from sampled data is relatively straightforward where categories are of known size and the number of categories~$|\mathcal{C}|$ is fairly small (so that a random sample provides large numbers of vertex pairs in each pair of categories), 
it is much more difficult when $|\mathcal{C}|$ is large and category sizes are not known. 
Our techniques thus extend the prior literature on block models and mixing rates to cases such as group interaction in OSNs and other large-scale social networks, in which one must estimate interaction among many groups of uncertain size from (typically non-uniformly) sampled data.  Our work also differs from much recent social network literature in being design-based rather than model-based; design-based inference is frequently easier to employ than model-based inference, although both approaches have merits~\cite{thompson:bk:2002}. %

\smallskip\noindent\textbf{Facebook colleges.}
The Facebook social graph has been measured and studied in the past. 
For example, \cite{Golder07_Rhythms_of_Social_Interaction}  studies the interactions between all 4.2M Facebook users in 492 universities in North America between Feb 2004 and March 2006. (As a side note, the interpretation is hindered by the full anonymization of user and universities.)
\cite{Traud2011} studies the social structure within 100 Facebook college categories. 
Given the above full datasets, one could apply \eqn{eq:w_AB_basic2} and create the category graph. 
In contrast, our methodological contribution lies in estimating the category graph from a sample of nodes, not from the fully known user graph. %

\newpage
\smallskip\noindent\textbf{Social graph visualization.} 
There exist many tools that visualize social graphs (including Facebook), for example~\cite{vizster_05,touchgraph,myfriendmap}. 
\url{www.geosocialmap.com} differs from most of these tools in that it 
(i)~is category-centric (vs user-centric), 
(ii)~contains an aggregated information view of \emph{entire} Facebook population, 
(iii)~is well suited for data exploration (\eg allows arbitrary selection of categories), and
(iv)~accepts as input any weighted graph with arbitrary set of node/edge attributes (ongoing work). 

\section{Conclusion}

\smallskip\noindent\textbf{Estimation performance.}  In this paper, we derive a number of category graph estimators for probability samples of nodes, uniform (\Sec{sec:Uniform sampling}) and non-uniform (\Sec{sec:Weighted sampling}). 
We evaluate their performance  
in simulation (\Sec{sec:Simulation Results}) and on Facebook  samples (\Sec{sec:Applications}). 
We showed 
that they all converge to their true values for reasonable sample sizes, a result we extend formally in the Appendix. 
Based on our evaluation, we also provide recommendations, summarized as follows.
When estimating category sizes, there is no universal choice between induced and star sampling. 
For example, the performance of the star estimator improves  (i)~in dense graphs, 
(ii)~in graphs with homogeneous node degree distribution,  (iii)~in graphs with weaker community structure, 
and (iv)~under sampling techniques that oversample high degree nodes.
 In contrast, when estimating the category edge weights, the star estimators are a clear winner; 
the induced subgraph estimators often need 5-10 times more samples to achieve the same accuracy. 
Finally, the sampling techniques strongly affect estimator efficiency. 
They can be ordered from best to worst as follows: UIS, \heur, RW and MHRW. 

\smallskip\noindent\textbf{Potential applications. }
We applied our methodology to samples of Facebook users and we estimated potentially interesting category graphs, such as the global friendship map, or the friendship network of US colleges.  
We visualized and made publicly available these weighted topologies at \url{www.geosocialmap.com}. 

In addition to  purely descriptive uses, the techniques described here can also be employed as a first step towards model-based analysis.  Using the unnormalized edge weights %
together with the number of possible edges within each cut yields the numbers of possible and realized edges needed for likelihood-based analysis of interaction probabilities.  For instance, given additional features associated with each category (\eg for universities, their size, location, ranking, and expense), one can model the inter-category mixing rates as a function of category features (\eg the effect of geographical distance on tie probability).  This permits both hypothesis testing for putative theories of tie formation and ex ante prediction of interaction rates among new or unobserved categories (given their hypothesized features) for extremely large, incompletely observed networks.  Given the large and growing literature on statistical modeling of networks (\eg \cite{pattison.et.al:jmp:2000,hoff.et.al:jasa:2002,vanduijn.et.al:statn:2004,hunter.handcock:jcgs:2006,butts:sm:2007a,robins.morris:sn:2007,goodreau.et.al:d:2009} among many others), the potential for applications in this area is substantial.

\bibliographystyle{abbrv}
{
\small
\bibliography{OSN_Sampling,ctb_refs}
}

\section*{Appendix: Consistency of the estimators} \label{sub:Appendix}

A desirable property of a statistical estimator is that of \emph{consistency}. A statistical estimator ($X_n$) is a function of the sample size ($n=|S|$), and is said to be \emph{consistent} if it converges in probability  ($\stackrel{P}{\longrightarrow}$) to the true value of interest ($\theta$) \cite{severini05} (which also implies \emph{asymptotic unbiasedness}). Formally: If $X_n \stackrel{P}{\longrightarrow} \theta$, as $n\rightarrow \infty$, then $X_n$ is said to be consistent for $\theta$. To prove the consistency of the estimators in this paper we invoke two classic theorems in probability: (1) The Law of Large Numbers, and (2) Slutsky's Theorem\footnote{For more details about these two theorems see \cite{severini05}.}; which require the following assumptions: For the uniform case we need to assume that the mean and variance are finite ($\theta<\infty$; $\sigma^2<\infty$); for the non-uniform case we need to make an additional assumption on the sampling weights so as to guarantee the consistency of the Hansen-Hurwitz (HH) estimator, specifically that the sum of the weights be bounded ($\sum_{v\in V} \w(v) \leq c$)\footnote{There are some alternate assumptions on the weights that can be made to guarantee convergence.}. Both of these conditions are satisfied for finite graphs.

\smallskip\noindent\textbf{LLN and Slutsky's Theorem}
\begin{theorem}[\textbf{Law of Large Numbers (LLN)}] Let $X_1, X_2, \dots$ be i.i.d. random variables with $EX_i = \theta$ and Var~$X_i= \sigma^2 < \infty$. Then $\bar{X}_n = \frac{1}{n} \sum_{v \in S} X(v) \stackrel{P}{\longrightarrow}~\theta$.
\end{theorem}

\begin{theorem}
[\textbf{Slutsky's Theorem}] Let $X_n  \stackrel{P}{\longrightarrow} \alpha$ and $Y_n  \stackrel{P}{\longrightarrow} \beta$, where $\alpha$ and $\beta$, respectively, are real numbers. Then

\begin{description}
\item[(p.1)] $ X_n + Y_n \stackrel{P}{\longrightarrow} \alpha + \beta $;
\item[(p.2)] $ X_n \cdot Y_n \stackrel{P}{\longrightarrow} \alpha \cdot \beta$;
\item[(p.3)] $ \frac{X_n}{Y_n} \stackrel{P}{\longrightarrow} \frac{\alpha}{\beta}$, where $\beta\neq 0$.
\end{description}
\end{theorem}

\smallskip\noindent\textbf{Uniform sampling estimators}
\begin{description}

\item[\eqn{eq:category_size_induced}:]  $|\est{A}|\ =\ N\cdot\frac{|S_A|}{|S|} = \frac{1}{n} \sum_{v \in S} 1_{\{ v \in A\}} \stackrel{P}{\longrightarrow} |A|$ by the LLN. 

\item[\eqn{eq:av node degree}:] $\est{k}_V = \frac{\sum_{v \in S} \deg(v)}{|S|} \stackrel{P}{\longrightarrow} k_V$ and 

$\est{k}_A = \frac{ \sum_{v \in S_A} \deg(v)}{|S_A|}\stackrel{P}{\longrightarrow}k_A$ by the LLN (as above).

\item[\eqn{eq:vol c 2 UIS}:] $\est{f}^\sss{vol}_A \ \ =\  \ \frac{1}{\vol(S)}  \sum_{s\in S}\sum_{v \in \mathcal{N}(s)}\!\! 1_{\{v\in A\}}$ 

$= \  \ \frac{\frac{1}{n}\sum_{s\in S}\sum_{v \in \mathcal{N}(s)}\!\! 1_{\{v\in A\}}}{\frac{1}{n}\vol(S)} \stackrel{P}{\longrightarrow} f^\sss{vol}_A$ by an application of the LLN to both the numerator and denominator, separately, followed by an application of Slutsky's Theorem (p.3).

\item[\eqn{eq:category_size_star}:] $|\est{A}|\ =\ N\cdot \est{f}^\sss{vol}_A \cdot \frac{\est{k}_V}{\est{k}_A}  \stackrel{P}{\longrightarrow} |A|$ by two applications of Slutsky's Theorem (p.2 and p.3) and consistency of the individual estimators.

\item[\eqn{eq:w_AB_Both end-nodes sampled}:] $\est{\w}(A,B)\ =\ \frac{ \sum_{a\in S_A}\sum_{b\in S_B} 1_{\{\{a,b\}\in E\}} }{|S_A|\cdot|S_B|}$

$= \ \frac{ \frac{N^2}{n^2}   \sum_{a\in S_A}\sum_{b\in S_B} 1_{\{\{a,b\}\in E\}} }{ \frac{N^2}{n^2} \sum_{a \in S} \sum_{b \in S} 1_{\{a \in S_A \textrm{ and } b \in S_B\}}} \stackrel{P}{\longrightarrow} \frac{|E_{A,B}|}{|A||B|}=\w(A,B)$ by LLN and Slutsky's Theorem (p.3).

\item[\eqn{eq:w_AB_star}:] $\est{\w}(A,B)\ =\ \frac{ \sum_{a\in S_A} |E_{a,B}| \ +\  \sum_{b\in S_B} |E_{b,A}|}{ |S_A|\cdot|\est{B}|\ +\ |S_B|\cdot|\est{A}|}=$

$\frac{  \frac{N}{n}  \sum_{a\in S_A} |E_{a,B}| \ +\   \frac{N}{n} \sum_{b\in S_B} |E_{b,A}|}{  \frac{N}{n} |S_A|\cdot|\est{B}|\ +\  \frac{N}{n} |S_B|\cdot|\est{A}| } \stackrel{P}{\longrightarrow} \frac{|E_{A,B}|+|E_{A,B}|}{|A||B|+|A||B|} = \w(A,B)$ by the LLN on numerator and denominator and then by five applications of Slutsky's Theorem (p.1, p.2, and p.3).
\end{description}

\smallskip\noindent\textbf{Non-uniform sampling estimators}

\begin{description}

\item[\eqn{f_tot}:] $\hat{x}_\sss{tot} = \frac{1}{n}\sum_{v\in S} \frac{x(v)}{\pi(v)}$ is shown to be consistent in \cite{HansenHurwitz1943}.

\item[\eqn{eq:category_size_induced_weighted}:] $ |\est{A}| \ = \  N\cdot\frac{\frac{1}{n}\w_\inv(S_A)}{\frac{1}{n}\w_\inv(S)}  \stackrel{P}{\longrightarrow} |A|$ by the consistency of the HH estimator in the numerator and denominator and then by Slutsky's Theorem (p.3).

\item[\eqn{eq:w_degree}:] $\est{k}_V = \frac{\frac{1}{n} \sum_{v \in S} \frac{ \deg(v)}{\w(v)}}{ \frac{1}{n}  \w_\inv(S)}  \stackrel{P}{\longrightarrow} k_V$ and 

$\est{k}_A = \frac{\frac{1}{n}  \sum_{v \in S_A} \frac{ \deg(v)}{\w(v)}}{\frac{1}{n}  \w_\inv(S_A)}  \stackrel{P}{\longrightarrow} k_A$ by the consistency of the HH estimator and Slutsky's Theorem (p.3).

\item[\eqn{vol c 2 WIS}:] $\est{f}^\sss{vol}_A\ \ =\  \ \frac{ \frac{1}{n}
\sum_{s\in S} \left(\frac{1}{\w(s)}\sum_{v \in \mathcal{N}(s)}\!\! 1_{\{v\in A\}}\right)
}{\frac{1}{n}\sum_{s\in S} \frac{\deg(s)}{\w(s)}}  \stackrel{P}{\longrightarrow} f^\sss{vol}_A$ by the consistency of the HH estimator and Slutsky's Theorem (p.3).

\item[\eqn{eq:category_size_star_WIS}:] $|\est{A}|\ =\ N\cdot \est{f}^\sss{vol}_A \cdot \frac{\est{k}_V}{\est{k}_A}  \stackrel{P}{\longrightarrow} |A|$ by the consistency of the estimators and Slutsky's Theorem (p.2 and p.3).

\item[\eqn{eq:w_AB_Both end-nodes sampled WIS}:] $\est{\w}(A,B)\ =\ \frac{ \frac{1}{n^2}  \sum_{a\in S_A}\sum_{b\in S_B} \frac{1_{\{\{a,b\}\in E\}}}{ \w(a)\cdot\w(b)} }{\frac{1}{n^2} \w_\inv(S_A)\cdot\w_\inv(S_B)}\\\stackrel{P}{\longrightarrow} \frac{|E_{A,B}|}{|A||B|}=\w(A,B)$ by the consistency of the HH estimator and Slutsky's Theorem (p.2 and p.3).

\item[\eqn{eq:w_AB_star_WIS}:] $\est{\w}(A,B)\ =\ \frac{\frac{1}{n} \sum_{a\in S_A} \frac{|E_{a,B}|}{\w(a)} \ +\  \frac{1}{n} \sum_{b\in S_B} \frac{|E_{b,A}|}{\w(b)} }{\frac{1}{n}  \w_\inv(S_A)\cdot|\est{B}|\ +\ \frac{1}{n} \w_\inv(S_B)\cdot|\est{A}|}=$

$\stackrel{P}{\longrightarrow} \frac{|E_{A,B}|+|E_{A,B}|}{|A||B|+|A||B|} =\w(A,B)$ by the consistency of HH estimators in the numerator and denominator and then by five applications of Slutsky's Theorem (p.1, p.2 and p.3).

\end{description}

\smallskip\noindent\textbf{A note on dependent samples}

These results continue hold in the case of dependent (correlated) samples, such as RW, under the condition that these samples converge asymptotically to UIS or WIS limits.  This follows from the ergodic theorem, which provides a corresponding LLN for convergent Markov Chains. For more technical details on the LLN in the context of dependent samples see \cite{Ribeiro2010}.

\end{document}